\documentclass[PRC,aps,showpacs,twocolumn,superscriptaddress,floatfix]{revtex4-1}

\usepackage{ulem}

\usepackage[dvipdfmx]{graphicx}
\usepackage{here}
\usepackage{amsmath,amssymb,times}
\usepackage{color}

\newcommand{\beq}{\begin{equation}}
\newcommand{\eeq}{\end{equation}}
\newcommand{\bea}{\begin{eqnarray}}
\newcommand{\eea}{\end{eqnarray}}

\begin{document}

\title{Neutron skin of $^{48}$Ca consistent with experimental data on skins
}

\author{Shingo~Tagami}
\affiliation{Department of Physics, Kyushu University, Fukuoka 812-8581, Japan}

\author{Jun~Matsui}
\affiliation{Department of Physics, Kyushu University, Fukuoka 812-8581, Japan}

\author{Maya~Takechi}
\affiliation{Niigata University, Niigata 950-2181, Japan}

\author{Masanobu~Yahiro}
\email[]{orion093g@gmail.com}
\affiliation{Department of Physics, Kyushu University, Fukuoka 812-8581, Japan}

\begin{abstract}
\noindent 
{\bf Background:}
In our previous paper, we predicted neutron skin $r_{\rm skin}$ and  proton, neutron, matter radii, $r_{\rm p}$, 
$r_{\rm n}$, $r_{\rm m}$  for $^{40-60,62,64}$Ca after determining the neutron dripline, 
using the Gogny-D1S Hartree-Fock-Bogoliubov (GHFB) with and without the angular momentum projection 
(AMP).  We found that effects of the AMP are small.  
Very lately, Tanaka {\it et al.} measured interaction cross sections $\sigma_{\rm I}$ for $^{42-51}$Ca, 
determined  matter radii $r_{\rm m}(\sigma_{\rm I})$ from the $\sigma_{\rm I}$, and deduced skin $r_{\rm skin}(\sigma_{\rm I})$ and $r_{\rm n}(\sigma_{\rm I})$ 
from the  $r_{\rm m}(\sigma_{\rm I})$ and the  $r_{\rm p}(\rm {exp})$ evaluated from the electron scattering. 
Comparing our results with the data, we find for $^{42-48}$Ca that  GHFB and GHFB+AMP reproduce 
$r_{\rm skin}(\sigma_{\rm I})$, $r_{\rm n}(\sigma_{\rm I})$, $r_{\rm m}(\sigma_{\rm I})$, but not for 
$r_{\rm p}(\rm {exp})$. 
 \\
{\bf Aim:} 
Our purpose is to determine a value of $r_{\rm skin}^{48}$ by using GHFB+AMP and 
the constrained GHFB  (cGHFB) in which the calculated value is fitted to $r_{\rm p}(\rm {exp})$. 
\\
{\bf Results:} 
For $^{42,44,46,48}$Ca, cGHFB hardly changes $r_{\rm skin}$, $r_{\rm m}$, $r_{\rm n}$
calculated with GHFB+AMP, except for $r_{\rm skin}^{48}$. 
For $r_{\rm skin}^{48}$,  the cGHFB result is $r_{\rm skin}^{48}=0.190$~fm, 
while  $r_{\rm skin}^{48}=0.159$~fm for GHFB+AMP. 
We should take the upper and the lower bound of GHFB+AMP and cGHFB.   
The result  $r_{\rm skin}^{48}=0.159-0.190$~fm  consists with the $r_{\rm skin}^{48}(\sigma_{\rm I})$ and 
the data  $r_{\rm skin}^{48}(\rm $E1$pE)$ obtained from high-resolution $E1$ polarizability experiment ($E1$pE).
Using the  $r_{\rm skin}^{48}$-$r_{\rm skin}^{208}$ relation with strong correlation of 
Ref.~\cite{Tagami:2020shn}, 
we transform the data $r_{\rm skin}^{208}$ determined by PREX and $E1$pE to the corresponding values, 
$r_{\rm skin}^{48}(\rm tPREX)$ and $r_{\rm skin}^{48}(\rm t$E1$pE)$, where the symbol `t' stands 
for the transformed data.  
Our result is consistent also for $r_{\rm skin}^{48}(\rm tPREX)$ and $r_{\rm skin}^{48}(\rm t$E1$pE)$. 
Eventually, for $^{42,44,46,48}$Ca, cGHFB reproduces $r_{\rm skin}(\sigma_{\rm I})$, $r_{\rm m}(\sigma_{\rm I})$, 
$r_{\rm n}(\sigma_{\rm I})$,  $r_{\rm p}(\rm {exp})$, while GHFB+AMP 
does $r_{\rm skin}(\sigma_{\rm I})$, $r_{\rm m}(\sigma_{\rm I})$, $r_{\rm n}(\sigma_{\rm I})$. 
 \end{abstract}

\maketitle

\section{Introduction and conclusion}
\label{Sec:Introduction}

{\it Background on experiments.} 
Neutron skin thickness $r_{\rm skin}$ is strongly correlated with the sloop parameter $L$ 
in the symmetric energy of nuclear matter~\cite{RocaMaza:2011pm,Brown:2013mga,Tagami:2020shn}. 
The $r_{\rm skin}$ is thus important to determine the EoS.

Horowitz, Pollock and Souder proposed a direct measurement 
for $r_{\rm skin}=r_{\rm n}-r_{\rm p}$~\cite{Hor01a}, where $r_{\rm p}$ and $r_{\rm n}$ are 
proton and neutron radii, respectively.  
The measurement consists of parity-violating and elastic electron scattering. The $r_{\rm n}$ is 
determined from the former experiment, and the $r_{\rm p}$ is from the latter.  
For $r_{\rm skin}^{208}$, in fact, 
the Lead Radius EXperiment (PREX)~\cite{PREX05,Abrahamyan:2012gp,Ong:2010gf}
yields   
\bea
r_{\rm skin}^{208}({\rm PREX}) =0.33^{+0.16}_{-0.18}=0.15-0.49~{\rm fm} .
\label{Eq:direct constraint}
\eea
The result has a large error. For this reason, the PREX-II and 
the $^{48}$Ca Radius EXperiment (CREX) are ongoing at Jefferson Lab~\cite{PREX05}. 

As an indirect measurement on $r_{\rm skin}$, the high-resolution $E1$ polarizability 
experiment ($E1$pE)  was made 
for $^{208}$Pb~\cite{Tamii:2011pv} and $^{48}$Ca~\cite{Birkhan:2016qkr} in RCNP. 
The results are
\bea
r_{\rm skin}^{208}(E1{\rm pE}) &=&0.156^{+0.025}_{-0.021}=0.135-0.181~{\rm fm}, 
\label{Eq:skin-Pb208-E1}
\\
r_{\rm skin}^{48}(E1{\rm pE}) &=&0.14-0.20~{\rm fm}.  
\label{Eq:skin-Ca48-E1}
\eea
For $^{208}$Pb, the central value 0.156~fm of  the indirect measurement is much smaller than 
0.33~fm of the direct measurement. This is a problem to be solved.  

Very lately, Tanaka~{\it et al.} published data on interaction cross sections 
$\sigma_{\rm I}$  for $^{42-51}$Ca~\cite{Tanaka:2019pdo}. 
The data have high accuracy, since the average error is 1.05\%. 
They determined $r_{\rm m}(\sigma_{\rm I}) $ from the $\sigma_{\rm I}$ 
using the Glauber  model, and deduced $r_{\rm skin}(\sigma_{\rm I})$ and $r_{\rm n}(\sigma_{\rm I})$ from 
the $r_{\rm m}(\sigma_{\rm I})$ and the $r_{\rm p}({\rm exp})$ of Ref.~\cite{Angeli:2013epw}.
For $^{48}$Ca, a value of $r_{\rm skin}(\sigma_{\rm I})$ is 
\bea
r_{\rm skin}^{48}(\sigma_{\rm I})=0.086 - 0.206~ {\rm fm}. 
\label{Eq:skin-Ca48-sigma-I}
\eea

Using the  $r_{\rm skin}^{48}$-$r_{\rm skin}^{208}$ relation ~\cite{Tagami:2020shn} 
with high correlation coefficient of $R=0.99$, 
\bea
r_{\rm skin}^{48}=0.5547~r_{\rm skin}^{208}+0.0718, 
\label{Eq:skin-208-48-1}
\eea
we  transform $r_{\rm skin}^{208}({\rm PREX})$ and $r_{\rm skin}^{208}(E1{\rm pE})$ 
to the corresponding values $r_{\rm skin}^{48}({\rm tPREX})$ and $r_{\rm skin}^{48}({\rm t}E1{\rm pE})$, 
where the symbol `t' stands for the transformed data.  
The transformed data are 
\bea
r_{\rm skin}^{48}({\rm tPREX})=0.155-0.344~{\rm fm}
\label{Eq:skin-Ca48-PREX-transformed}
\eea
for PREX and 
\bea
r_{\rm skin}^{48}({\rm t}E1{\rm pE})=0.147-0.172~{\rm fm}
\label{Eq:skin-Ca48-E1-transformed}
\eea
for $E1$pE. 

{\it Background on theories:} 
As an {\it ab initio} method for Ca isotopes,
we should consider the coupled-cluster method~\cite{Hagen:2013nca,Hagen:2015yea} with chiral interaction. 
The coupled-cluster result~\cite{Hagen:2015yea} 
\bea
r_{\rm skin}^{48}({\rm CC})=0.12 - 0.15~ {\rm fm}
\label{Eq:skin-Ca48-CC}
\eea 
is consistent with data $r_{\rm skin}^{48}(E1{\rm pE})$ and $r_{\rm skin}^{48}(\sigma_{\rm I})$. 
 
\begin{figure}[H]
\centering
\vspace{0cm}
\includegraphics[width=0.4\textwidth]{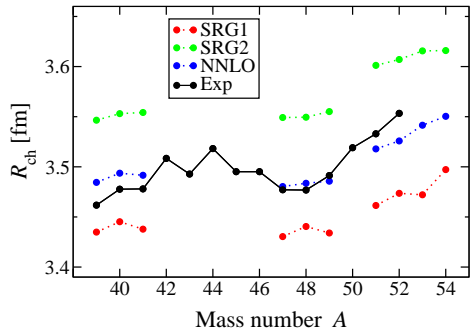}
\caption{
$A$ dependence of charge radii $R_{\rm ch}$ for $^{40-54}$Ca taken from Ref.~\cite{Ruiz:2016gne}. 
}
\label{fig:$A$ dependence of R-ch}
\end{figure}

Among effective interactions, NNLO$_{\rm sat}$~\cite{Ekstrom:2015rta} is 
a chiral interaction constrained by radii and binding energies of selected nuclei 
up to $A \approx 25$~\cite{Hagen:2015yea}, where $A$ is the mass number. 
In fact, the {\it ab initio} calculations were done 
for Ca isotopes~\cite{Ekstrom:2015rta, Hagen:2015yea, Ruiz:2016gne}.  
As shown in Fig.~\ref{fig:$A$ dependence of R-ch}, 
Garcia Ruiz {\it et. al.} evaluated the charge radii $R_{\rm ch}$ for 
$^{39-54}$Ca~\cite{Ruiz:2016gne}, 
using the coupled-cluster method with two low-momentum effective interactions, 
SRG1 of Ref.~\cite{Hebeler:2010xb} and SRG2 of Ref.~\cite{Furnstahl:2013oba}, 
that are derived from the chiral  interaction with the renormalization group method. 
The SRG1 (SRG2) yields the lower (upper) bound of $r_{\rm ch}$. 
Difference of the two results is $\sim 1.2$~fm.

In our previous paper of Ref.~\cite{Tagami:2019svt}, we predicted 
$r_{\rm p}$, $r_{\rm n}$, matter radii $r_{\rm m}$,  $r_{\rm skin}$ for $^{40-60,62,64}$Ca after determining 
the neutron dripline, using the Gogny-D1S Hartree-Fock-Bogoliubov (GHFB) with 
the angular momentum projection (AMP)~\cite{Tagami-AMP} for even nuclei and GHFB for odd nuclei.  
For odd nuclei, GHFB+AMP calculations are not feasible. The reason is shown in Sec.~\ref{Sec:GFHB+AMP}. 
Using the Kyushu (chiral) $g$-matrix folding model~\cite{Toyokawa:2014yma,Toyokawa:2015zxa,Toyokawa:2017pdd}, we also predicted  reaction cross section  
$\sigma_{\rm R}$ for $^{40-60,62,64}$Ca  scattering on a $^{12}$C target at $280$~MeV/nucleon, since 
Tanaka {\it el al.} measured interaction cross sections $\sigma_{\rm I} (\approx \sigma_{\rm R})$ 
for $^{42-51}$Ca. 

In our previous paper, we first confirmed that effects of the AMP are small for even nuclei. 
GHFB and GHFB+AMP reproduce the one-neutron separation energy $S_{1}$ and 
the two-neutron separation energy $S_{2}$ in $^{41-58}$Ca~\cite{HP:NuDat 2.7,Tarasov2018,Michimasa2018obr}. 
Using $S_{1}$ and $S_{2}$, we found 
that $^{64}$Ca is an even-dripline nucleus and $^{59}$Ca is an odd-dripline nucleus. 
As for $E_{\rm B}$, our results are consistent with the data~\cite{HP:NuDat 2.7} in $^{40-58}$Ca.    
Comparing our results with new data of Ref.~\cite{Tanaka:2019pdo}, we find for $^{42-48}$Ca that  
GHFB and GHFB+AMP reproduce 
$r_{\rm skin}(\sigma_{\rm I})$, $r_{\rm n}(\sigma_{\rm I})$, $r_{\rm m}(\sigma_{\rm I})$, but not for 
$r_{\rm p}(\rm {exp})$.

{\it Aim and conclusion:}
In this paper, we determine a value of $r_{\rm skin}^{48}$ with GHFB+AMP and 
the constrained GHFB  (cGHFB) in which the calculated value is fitted to $r_{\rm p}(\rm {exp})$. 
For $r_{\rm skin}^{48}$,  the cGHFB result is $r_{\rm skin}^{48}=0.19$~fm,  while  $r_{\rm skin}^{48}=0.159$~fm for GHFB+AMP. 
We should take the upper and the lower bound of GHFB+AMP and cGHFB.  
Our result is 
\bea
r_{\rm skin}^{48}=0.159-0.19~{\rm fm}.
\label{eq-skin-48-TW}
\eea

\begin{figure}[H]
\centering
\vspace{0cm}
\includegraphics[width=0.4\textwidth]{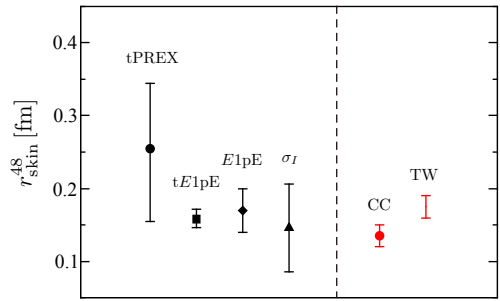}
\caption{
Comparison among four data and two theoretical results of coupled-cluster theory 
and this work (TW) for $r_{\rm skin}^{48}$; namely, 
$r_{\rm skin}^{48}({\rm tPREX})$, $r_{\rm skin}^{48}({\rm t}E1{\rm pE})$, 
$r_{\rm skin}^{48}(E1{\rm pE})$, $r_{\rm skin}^{48}(\sigma_I)$,  
$r_{\rm skin}^{48}({\rm CC})$, $r_{\rm skin}^{48}({\rm TW})$. 
}
\label{fig:fig2}
\end{figure}

Our result is consistent with $r_{\rm skin}^{48}(\rm $E1$pE)$, $r_{\rm skin}^{48}(\sigma_{\rm I})$, 
$r_{\rm skin}^{48}(\rm tPREX)$ and $r_{\rm skin}^{48}(\rm t$E1$pE)$, as shown in Fig.~\ref{fig:fig2}. 
The figure also shows that our result is also consistent 
with the coupled-cluster one of  Eq.~\eqref{Eq:skin-Ca48-CC}.

We recapitulate  our models in Sec.~\ref{Sec:Theoretical framework}, 
and show our results for neighbor nuclei of  $^{48}$Ca 
in Sec.~\ref{Sec:Results for neighbor nuclei of  $^{48}$Ca}.

\section{Models}
\label{Sec:Theoretical framework}
 
We recapitulate GHFB, GFHB+AMP and cGHFB.

\subsection{GFHB+AMP}
\label{Sec:GFHB+AMP}

In GHFB+AMP, the total wave function  $| \Psi^I_{M} \rangle$ with the AMP is defined by 
\begin{equation}
 | \Psi^I_{M} \rangle =
 \sum_{K, n=0}^{N} g^I_{K n} \hat P^I_{MK}|\Phi_n \rangle ,
\label{eq:prjc}
\end{equation}
where $\hat P^I_{MK}$ is the angular-momentum-projector and the 
$|\Phi_n \rangle$ for $n=0,1,\cdots,N$ are mean-field (GHFB) states, 
where $N+1$ is the number  of the  states.  
The coefficients $g^I_{K n}$ are obtained by solving the Hill-Wheeler equation
\begin{equation}
 \sum_{K^\prime n^\prime }{\cal H}^I_{Kn,K^\prime n^\prime }\ g^I_{K^\prime n^\prime } =
 E_I\,
 \sum_{K^\prime n^\prime }{\cal N}^I_{Kn,K^\prime n^\prime }\ g^I_{K^\prime n^\prime },
\end{equation}
with the Hamiltonian and norm kernels defined by
\begin{equation}
 \left\{ \begin{array}{c}
   {\cal H}^I_{Kn,K^\prime n^\prime } \\ {\cal N}^I_{Kn,K^\prime n^\prime } \end{array}
 \right\} = \langle \Phi_n |
 \left\{ \begin{array}{c}
   \hat{H} \\ 1 \end{array}
 \right\} \hat{P}_{KK^\prime }^I | \Phi_{n'} \rangle.
\end{equation}

For even nuclei, there is no blocking state, i.e., $N=0$ in the Hill-Wheeler equation. 
We can thus perform  GHFB+AMP. 
However,  we have to find the value of $\beta$ at 
which the ground-state energy becomes minimum.  
In this step, the AMP has to be performed for any $\beta$, so that the Hill-Wheeler calculation is still heavy. 
In fact, the AMP is not taken for mean field calculations in many works; see for example Ref.~\cite{HP:AMEDEE}.  
The reason why we do not take into account $\gamma$ deformation is 
that the deformation does not affect  $\sigma_{\rm R}$~\cite{Sumi:2012fr}. 
As for GHFB, meanwhile, we do not have to solve the Hill-Wheeler equation.

For odd nuclei,  we must put a quasi-particle in a level. The number $N$ of the blocking states 
are very large. This makes it difficult to solve  the Hill-Wheeler equation. 
Furthermore, we have to confirm that the resulting  $| \Psi^I_{M} \rangle$ converges as $N$ goes up 
for any set of two deformations  $\beta$ and $\gamma$. This procedure is extremely time-consuming. 
For this reason, we do not consider the AMP for  odd nuclei. 
As for GHFB, meanwhile, we do not have to solve the Hill-Wheeler equation, since 
we consider the one-quasiparticle state with the lowest energy. 
However, it is not easy  to find the values of 
$\beta$ and $\gamma$ at which the energy becomes minimum in the $\beta$-$\gamma$ plane.

As a result of the heavy calculations for even nuclei, 
we find that $\beta$ is small for GHFB+AMP; see the table I of Ref.~\cite{Tagami:2019svt} for the values 
of $\beta$. 
Meanwhile, the mean-field (GHFB) calculations yield that the energy surface becomes minimum 
at $\beta=0$. The fact that $\beta=0$ for GHFB and small for GHFB+AMP 
yields small difference between GHFB results and GHFB+AMP ones.  
Therefore, we consider GHFB+AMP for even nuclei and GHFB for odd nuclei.

\subsection{Constrained-GFHB}
\label{Sec:Constrained-GFHB}

The difference between $r_p({\rm GHFB+AMP})$ 
and $r_p({\rm exp})$ is largest for $^{48}$Ca in $^{40-52}$Ca. 
In order to fit $r_p$ to the central value of $r_p({\rm exp})$, one use constrained HFB; 
for example, see Ref.~\cite{Khan:2010mv}. 
In the framework of GHFB, we modify the Hamiltonian as  
\bea
\hat H_{\rm constraint} \equiv \hat H  +\lambda \hat Q 
\eea
with 
\bea
\hat Q=\hat r_p^2-[r_p({\rm exp})]^2,
\eea 
and take the expectation value $\langle \lambda |\hat Q | \lambda \rangle$ 
of  $\hat Q$ with the constrained-GHFB (cGHFB) solution $| \lambda \rangle$. 
The $r_p({\rm  cGHFB})$ determined by cGHFB
agrees with $r_p({\rm exp})$ under the condition 
\bea
\frac{d \langle \lambda |\hat Q | \lambda \rangle}{d \lambda}|_{\lambda=0}=0. 
\eea
In actual calculations,  we use the augmented Lagrangian method~\cite{Staszczak:2010zt}.

\section{Results for neighbor nuclei of  $^{48}$Ca}
\label{Sec:Results for neighbor nuclei of  $^{48}$Ca}

As neighbor nuclei of  $^{48}$Ca, we consider $^{42-51}$Ca, since the data are  available 
for $r_{\rm skin}$, $r_{\rm m}$, $r_{\rm n}$, $r_{\rm p}$.

\begin{figure}[H]
\centering
\vspace{0cm}
\includegraphics[width=0.4\textwidth]{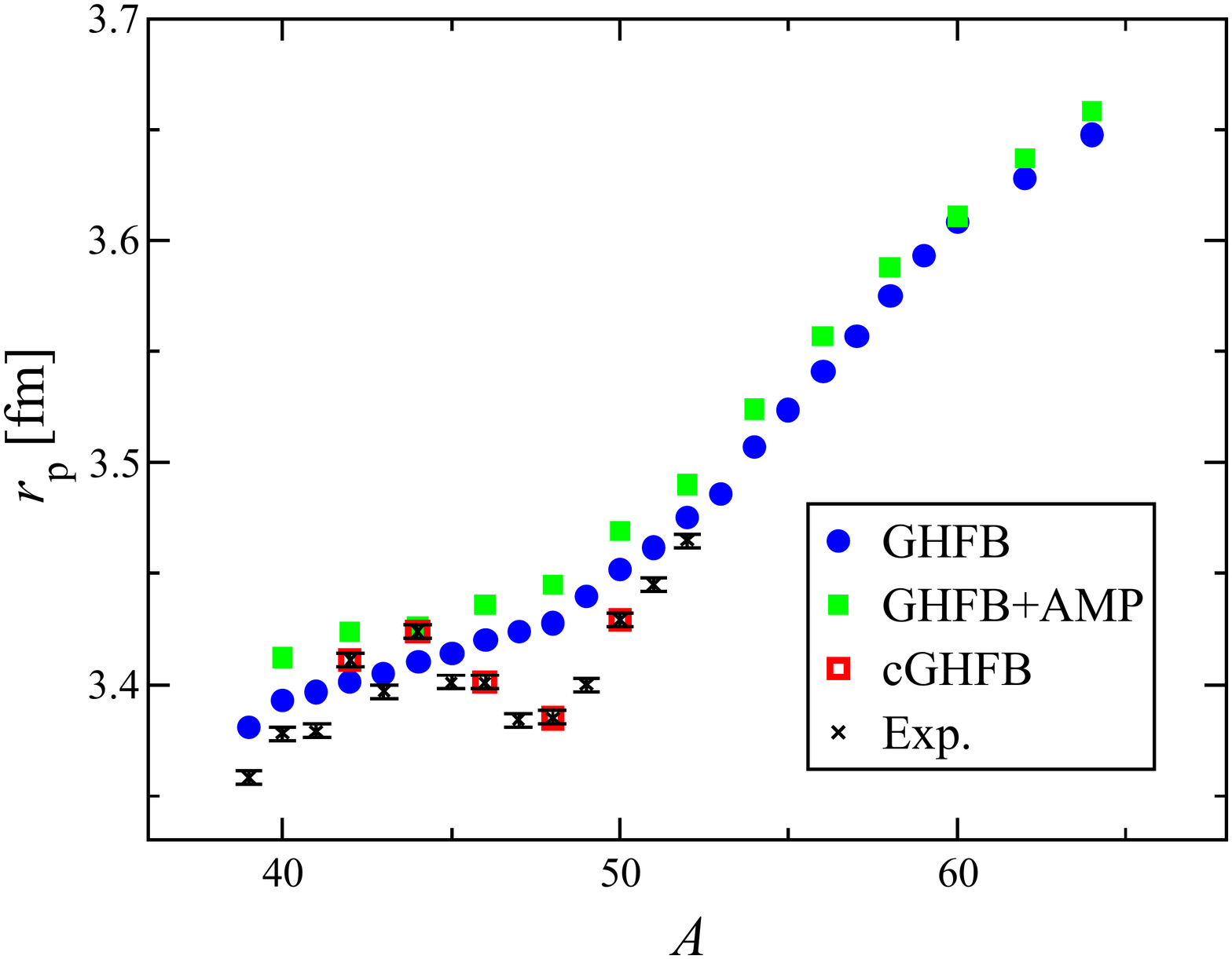}
\includegraphics[width=0.4\textwidth]{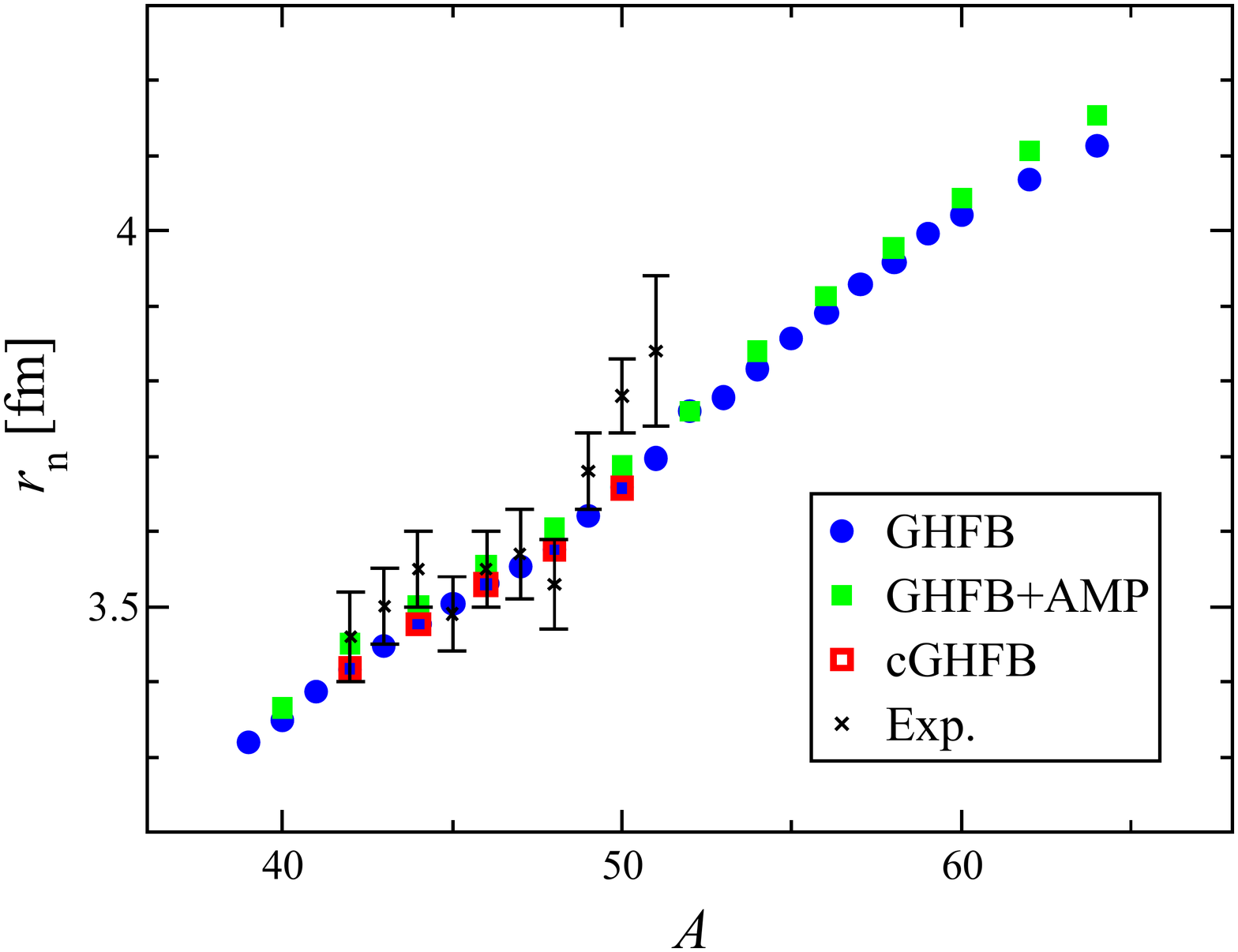}
\includegraphics[width=0.4\textwidth]{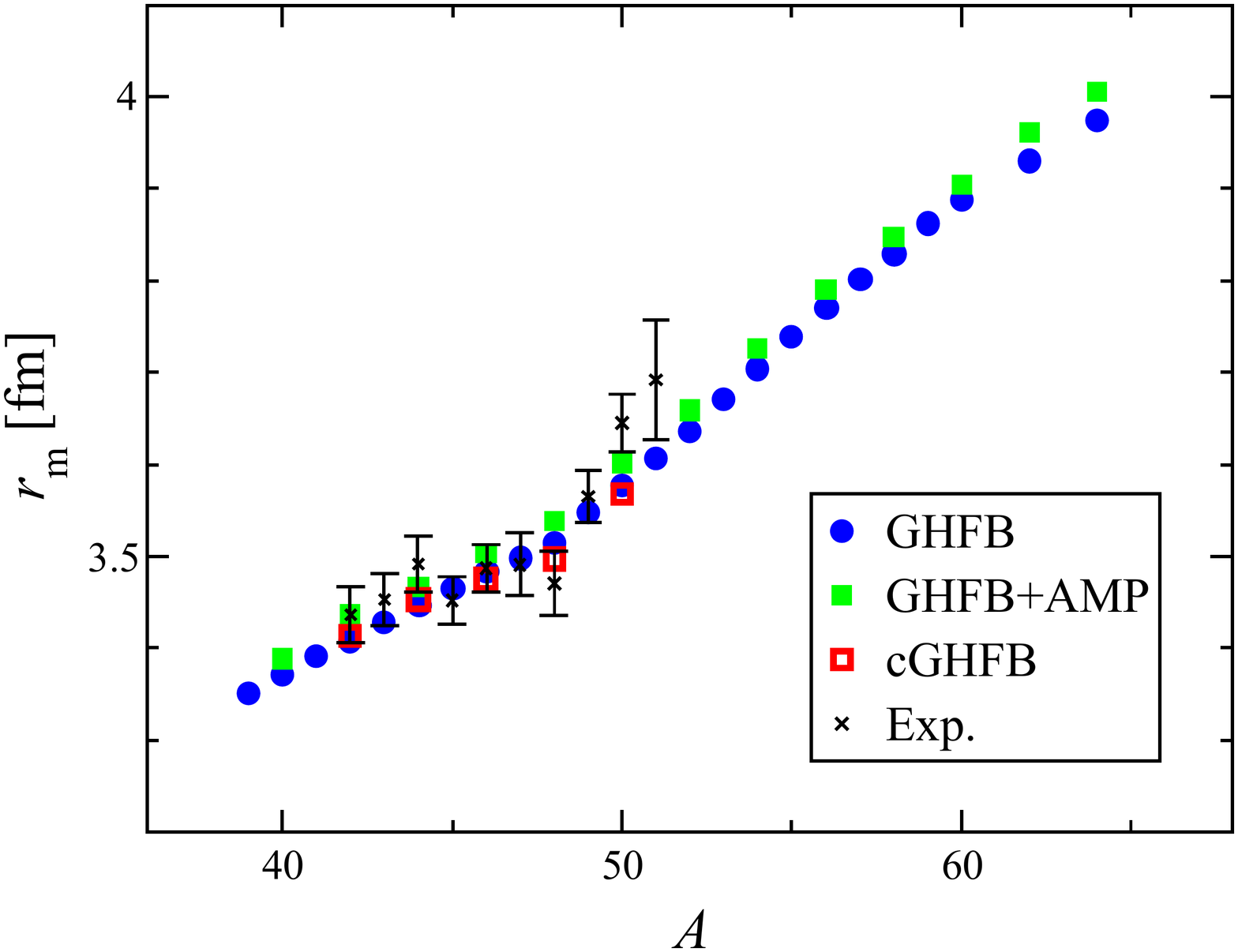}
\includegraphics[width=0.4\textwidth]{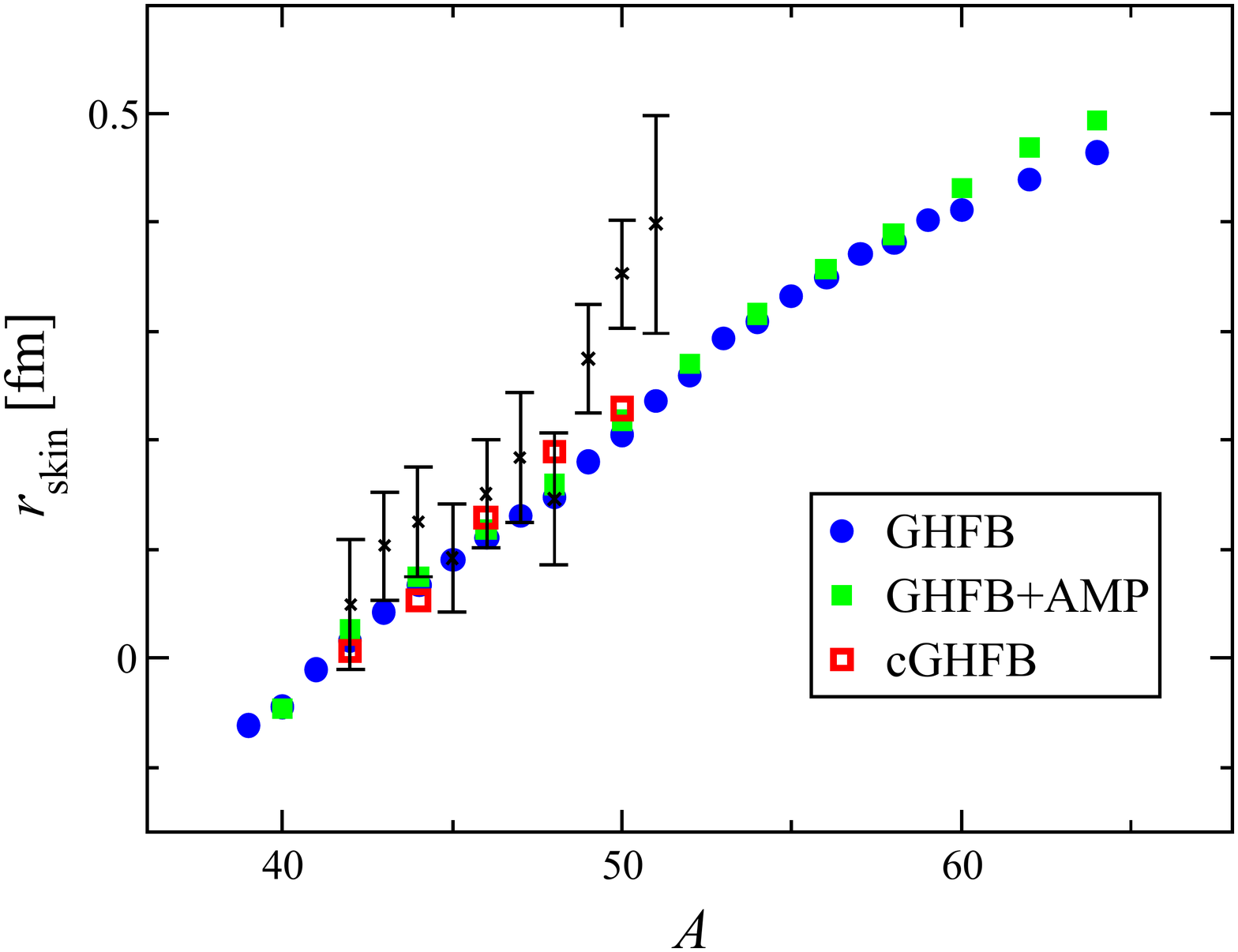}
\vspace{-10pt}
\caption{$A$ dependence of  $r_{\rm p}$, $r_{\rm n}$, $r_{\rm m}$, $r_{\rm skin}$. 
Closed  circles denote the GHFB results, while  closed squares correspond 
to the GHFB+AMP results.  
Open squares show the results of cGHFB for $^{42,44,46,48,50}$Ca.
Experimental data are taken from Refs.~\cite{Angeli:2013epw,Tanaka:2019pdo}. 
}
\label{fig:A dependence of radii}
\end{figure}

Figure \ref{fig:A dependence of radii} shows $r_{\rm p}$, $r_{\rm n}$, $r_{\rm m}$, $r_{\rm skin}$ 
as a function of $A$. 
As for $r_{\rm p}$, the results of GHFB+AMP and GHFB do not reproduce 
the data~\cite{Angeli:2013epw} for $^{39-52}$Ca. 
For $^{42,44,46,48,50}$Ca, we then do cGHFB calculations to fit the theoretical value 
to the central value of $r_p({\rm exp})$. 

The cGFHB results hardly change the values of 
$r_{\rm n}$, $r_{\rm m}$, $r_{\rm skin}$ except for $r_{\rm skin}^{48}$; see Table~\ref{Radii of cGHFB} for 
the numerical values of  cGHFB and  Table~\ref{Radii for Ca isotopes. } for the numerical values of  
GHFB and GHFB+AMP. 
The deviation for $r_{\rm p}$ is thus not  important except for $r_{\rm skin}^{48}$. 
We then take the lower and the upper bound of GHFB+AMP and cGHFB, as shown in 
our conclusion in Sec.~\ref{Sec:Introduction}.

As for  $r_{\rm skin}$, 
the results of GHFB, GHFB+AMP, cGHFB  reproduce 
the data~\cite{Tanaka:2019pdo} for $^{42-48}$Ca, but underestimate the data for $^{49-51}$Ca. 
The difference between GHFB+AMP and GHFB  is small for even Ca isotopes, indicating that 
effects of AMP are small. 
Eventually,  for $^{42,44,46,48}$Ca, cGHFB reproduce $r_{\rm skin}(\sigma_{\rm I})$, $r_{\rm m}(\sigma_{\rm I})$, 
$r_{\rm n}(\sigma_{\rm I})$,  $r_{\rm p}(\rm {exp})$, while GHFB+AMP 
does $r_{\rm skin}(\sigma_{\rm I})$, $r_{\rm m}(\sigma_{\rm I})$, $r_{\rm n}(\sigma_{\rm I})$.

The data on $r_m$  has a kink at $A=48$. 
Qualitatively, $r_m$ may be in inverse proportion to the binding energy per nucleon, $E_{\rm B}/A$.
We then consider a dimensionless quantity  $\alpha \equiv r_{m}E_{\rm B}/(A\hbar c)$, where 
the central values of data~\cite{Tanaka:2019pdo,HP:NuDat 2.7} are taken for  $r_{m}$ and  $E_{\rm B}/A$. 
The values of $\alpha$ are tabulated in Table~\ref{table-a}. 
The average of  $\alpha$ and its error  are 
\bea
\alpha=0.1535 (9) 
\label{Eq-a}
\eea
for  $^{42-51}$Ca, indicating that  $r_m$ is in inverse proportion to $E_{\rm B}/A$. 
We can say that the kink comes from the shell effect.

\begin{table}[htb]
\begin{center}
\caption
{Radii of constrained GHFB for $^{42,44,46,48,50}$Ca.
 }
 \begin{tabular}{ccccc}
  \hline
$A$  & $r_n$~fm & $r_p$~fm & $r_m$~fm & $r_{\rm skin}$~fm \\
  \hline
  42   & 3.417  & 3.411 &	3.414 &	0.006  \cr
  44   & 3.477 	&3.424 	&3.453 	&0.053   \cr
  46   & 3.530 &	3.401 	&3.475 	&0.129   \cr
  48   &3.575  &3.385  &3.497 & 0.190 \cr
  50   & 3.658 &	3.429 	&3.568 	&0.229  \cr
  \hline
 \end{tabular}   
 \label{Radii of cGHFB}
\end{center} \end{table}

\begin{table}[H]
\begin{center}
\caption
{Radii for Ca isotopes. 
The superscript ``AMP'' stands for the results of GHFB+AMP, and no
superscript corresponds to those of GHFB. 
 }
 \begin{tabular}{ccccccccc}
  \hline
$A$ & $r_n^{\rm AMP}$fm & $r_p^{\rm AMP}$fm & $r_m^{\rm AMP}$fm & $r_{\rm skin}^{\rm AMP}$fm &
$r_n$fm & $r_p$fm & $r_m$fm & $r_{\rm skin}$fm \\
  \hline
  39 & & & & &3.320  & 3.381  & 3.351  & -0.061  \cr
  40 & 3.366 & 3.412 & 3.389 & -0.046 & 3.349 & 3.393 & 3.371  & -0.044 \cr
  41 &  &  &  &  & 3.387 & 3.397 & 3.392 & -0.010  \cr
  42 & 3.451 & 3.424 & 3.438 & 0.026 & 3.417 & 3.401 & 3.409  & -0.010 \cr
  43 &  &  &  &  & 3.448 & 3.405 & 3.428 & 0.043  \cr
  44 & 3.501 & 3.426 & 3.467  & 0.075 & 3.477 & 3.410 & 3.447 &  0.067 \cr
  45 &  &  &  &  & 3.504 & 3.414 & 3.465 & 0.090  \cr
  46 & 3.555 & 3.436 & 3.504 & 0.118 & 3.530 & 3.420 & 3.483 &  0.110 \cr
  47 &  &  &  &  & 3.554 & 3.424 & 3.499 & 0.130  \cr
  48 & 3.604 & 3.445 & 3.539 & 0.159 & 3.576 & 3.428 & 3.515 & 0.148 \cr
  49 &  &  &  &  & 3.621 & 3.440 & 3.548 & 0.181  \cr
  50 & 3.687 & 3.469 & 3.601 & 0.218 & 3.658 & 3.452 & 3.577 &  0.206 \cr
  51 &  &  &  &  & 3.698 & 3.462 & 3.607 & 0.236 \cr
  52 & 3.760 & 3.490 & 3.659 & 0.270 & 3.734 & 3.475 & 3.659 &  0.270 \cr
  53 &  &  &  &  & 3.779 & 3.486 & 3.671 & 0.293 \cr
  54 & 3.840 & 3.524 & 3.726 & 0.316 & 3.817 & 3.507 & 3.705 &  0.310 \cr
  55 &  &  &  &  & 3.856 & 3.524 & 3.739 & 0.332  \cr
  56 & 3.913 & 3.557 & 3.790 & 0.357 & 3.891 & 3.541 & 3.770 &  0.350 \cr
  57 &  &  &  &  & 3.928 & 3.557 & 3.802 & 0.370  \cr
  58 & 3.977 & 3.588 & 3.847 & 0.389 & 3.958 & 3.575 & 3.830 &  0.383 \cr
  59 &  &  &  &  & 3.995 & 3.593 & 3.863 & 0.402  \cr
  60 & 4.043 & 3.611 & 3.904 & 0.432 & 4.020 & 3.608 & 3.888 &  0.412 \cr
  62 & 4.106 & 3.637 & 3.961 & 0.469 & 4.067 & 3.628 & 3.931 &  0.439 \cr
  64 & 4.153 & 3.658 & 4.005 & 0.494 & 4.113 & 3.648 & 3.974 &  0.465 \cr
  \hline
 \end{tabular}   
 \label{Radii for Ca isotopes. }
\end{center} \end{table}

\begin{table}[H]
\begin{center}
\caption
{Numerical values of  $r_{\rm m}(\sigma_{\rm I})$ for $^{42-51}$Ca. 
The $r_{\rm m}(\sigma_{\rm I})$ are taken from  Ref.~\cite{Tanaka:2019pdo}, 
and the data on $E_{\rm B}/A$ are  from Ref.~\cite{HP:NuDat 2.7}.
 }
\begin{tabular}{cccc}
\hline\hline
A & $r_{\rm m}({\sigma_I})$~fm &  $E_{\rm B}/A$~MeV &  $\alpha$ \\
\hline
42 & 3.437	&8.616563	&0.1501 \\
43	&3.453	&8.600663	&0.1505 \\
44	&3.492	&8.658175	&0.1532 \\
45	&3.452	&8.630545	&0.1510 \\
46	&3.487	&8.66898	      &0.1532 \\
47	&3.491	&8.63935	      &0.1528 \\
48	&3.471	&8.666686	&0.1524 \\
49	&3.565	&8.594844	&0.1553 \\
50	&3.645	&8.55016	      &0.1579 \\
51	&3.692	&8.476913	&0.1586 \\ 
\hline
\end{tabular}
 \label{table-a}
 \end{center} \end{table}


Table~\ref{Radii-exp} shows the experimental data~\cite{Tanaka:2019pdo} 
on $r_{\rm m}(\sigma_{\rm I})$, $r_{\rm n}(\sigma_{\rm I})$, 
$r_{\rm skin}(\sigma_{\rm I})$ for $^{42-51}$Ca, 
together with  $r_{\rm p}({\rm exp})$~\cite{Angeli:2013epw}.

\begin{table}[htb]
\begin{center}
\caption
{Numerical values of $r_{\rm p}({\rm exp})$, $r_{\rm m}(\sigma_{\rm I})$, $r_{\rm n}(\sigma_{\rm I})$, 
$r_{\rm skin}(\sigma_{\rm I})$ for $^{42-51}$Ca. 
The numerical values on $r_{\rm m}(\sigma_{\rm I})$, $r_{\rm n}(\sigma_{\rm I})$, 
$r_{\rm skin}(\sigma_{\rm I})$ are taken from  Ref.~\cite{Tanaka:2019pdo}, where the systematic error is included. 
The $r_{\rm p}({\rm exp})$ are deduced from the electron scattering~\cite{Angeli:2013epw}. 
Note that Tanaka {\it et al.} provide us the numerical values of  
$r_{\rm m}(\sigma_{\rm I})$, $r_{\rm skin}(\sigma_{\rm I})$, $r_{\rm n}(\sigma_{\rm I})$. 
 }
\begin{tabular}{ccccc}
\hline\hline
A & $r_{\rm p}({\rm exp})$~fm & $r_{\rm m}(\sigma_{\rm I})$~fm &  $r_{\rm n}(\sigma_{\rm I})$~fm & $r_{\rm skin}(\sigma_{\rm I})$~fm \\
\hline
42 & $3.411 \pm 0.003$ & $3.437 \pm 0.030$ & $3.46\pm 0.06$ & $0.049\pm 0.06$ \\
43 & 3.397 $\pm$ 0.003& 3.453 $\pm$ 0.029 & 3.50 $\pm$ 0.05 & 0.103 $\pm$ 0.05\\
44 & 3.424 $\pm$ 0.003 & 3.492 $\pm$ 0.030 & 3.55 $\pm$ 0.05 & 0.125 $\pm$ 0.05\\
45 & 3.401 $\pm$ 0.003 & 3.452 $\pm$ 0.026 & 3.49 $\pm$ 0.05 & 0.092 $\pm$ 0.05\\
46 & 3.401 $\pm$ 0.003 & 3.487 $\pm$ 0.026 & 3.55 $\pm$ 0.05 & 0.151 $\pm$ 0.05\\
47 & 3.384 $\pm$ 0.003 & 3.491 $\pm$ 0.034 & 3.57 $\pm$ 0.06  & 0.184 $\pm$ 0.06\\
48 & 3.385 $\pm$ 0.003 & 3.471 $\pm$ 0.035 & 3.53 $\pm$ 0.06 & 0.146 $\pm$ 0.06\\
49 & 3.400 $\pm$ 0.003 & 3.565 $\pm$ 0.028 & 3.68 $\pm$ 0.05 & 0.275 $\pm$ 0.05\\
50 & 3.429 $\pm$ 0.003 & 3.645 $\pm$ 0.031 & 3.78 $\pm$ 0.05  & 0.353 $\pm$ 0.05\\
51 & 3.445 $\pm$ 0.003 & 3.692 $\pm$ 0.066 & 3.84 $\pm$ 0.10 & 0.399 $\pm$ 0.10\\
\hline
\end{tabular}
 \label{Radii-exp}
 \end{center} \end{table}


\section*{Acknowledgements}
We thank Dr. Tanaka and Prof. Fukuda for providing the data on radii and interaction cross sections and 
helpful comments.  
M. Y. expresses our gratitude to  Dr. Y. R. Shimizu 
for his useful information.


\end{document}